\documentclass[12pt]{article}
\usepackage{graphicx}
\input epsf
\usepackage{amssymb}
\parskip=\medskipamount
\def\om{\omega}   
\def\ID{\relax{\rm l\kern-.18 em D}}
\def\IE{\relax{\rm l\kern-.18 em E}}
\def\IK{\relax{\rm l\kern-.18 em K}}
\def\IL{\relax{\rm I\kern-.18 em L}}
\def\IN{\relax{\rm I\kern-.18 em N}}
\def\IR{\relax{\rm I\kern-.18 em R}}
      
\def\uno{\relax{\rm 1\kern-.18 em l}}

\def\IK{\relax{\rm l\kern-.18 em K}}
\def\IL{\relax{\rm I\kern-.18 em L}}
\def\IN{{\Bbb N}}
\def\IR{{\Bbb  R}}
\def\ii{\rm i\,}

\def\Re{\mathop{\rm Re}\nolimits}
\def\Im{\mathop{\rm Im}\nolimits}

\def\smallonehalf{\frac{{}_1}{{}^2}}

\def\wt{\widetilde}
\def\frac#1#2{{#1\over #2}}

\def\ptos{\leaders\hbox to 2mm{\hfil{.}\hfil}\hfill}
\def\\{\hfill\break}

\def\<#1>{\langle#1\rangle}
\def\ii{{\rm i\,}}

\font\tenfrak=eufm10  \font\sevenfrak=eufm7  \font\fivefrak=eufm5
\newfam\frakfam
\textfont\frakfam=\tenfrak\scriptfont\frakfam=\sevenfrak
\scriptscriptfont\frakfam=\fivefrak

\font\tengoth=eufm10 scaled\magstep1 \font\sevengoth=eufm7
\font\fivegoth=eufm5
\newfam\gothfam
\textfont\gothfam=\tengoth\scriptfont\gothfam=\sevengoth
   \scriptscriptfont\gothfam=\fivegoth

\newtheorem{proposicion}{Proposition}

\begin{document}

\title{A new approach to the higher-order superintegrability  of 
the  Tremblay-Turbiner-Winternitz system }

\author{ Manuel F. Ra\~nada  \\ [3pt]
{\sl Dep. de F\'{\i}sica Te\'orica and IUMA } \\
  {\sl Universidad de Zaragoza, 50009 Zaragoza, Spain}     }
\date{ }
\maketitle

\begin{abstract}
The higher-order superintegrability of systems separable in polar coordinates is studied  using an approch that was previously applied for the study of the superintegrability of a generalized Smorodinsky-Winternitz system.
The idea is that the additional constant of motion can be factorized as the product of  powers  of two particular rather simple complex functions (here denoted by $M$ and $N$). 
This technique leads to a proof of the superintegrability of the 
TremblayÐTurbinerÐWinternitz system and to the explicit expression of the 
constants of motion.
A second family (related with the first one) of superintegrable systems is also studied. 
\end{abstract}

\begin{quote}
{\sl Keywords:}{\enskip} Integrability. Superintegrability. Nonlinear systems.
Higher-order  constants of motion. Complex factorization. 

{\sl Running title:}{\enskip}
Higher-order superintegrability of the TTW system.

AMS classification:  37J35 ; 70H06

PACS numbers:  02.30.Ik ; 05.45.-a ; 45.20.Jj
\end{quote}

\vfill
\footnoterule{\small
\begin{quote}
{\tt E-mail: {mfran@unizar.es}  }
\end{quote}
}

\newpage

\section{Introduction}

 A superintegrable system is a system that is integrable
(in the sense of Liouville-Arnold) and that, in addition to this,
possesses more constants of motion than degrees of freedom.  
Most of known superintegrable system are systems which admit 
separability in two (or more) different coordinate systems; for example, 
the so-called Smorodinsky-Winternitz (SW) potential 
$$
   V_{sw}  =  {\smallonehalf}\om_0^2(x^2 + y^2)
 + {k_2\over x^2}  + {k_3\over y^2} \,,
$$ 
is separable in Cartesian and polar coordinates \cite{FrMS65}--\cite{GroPoSi95}. 
A very important point is that the three constants of motion are quadratic in the
velocities (momenta). In fact, if we call superseparable a system
that admits Hamilton-Jacobi  separation of variables
(Schr\"odinger in the quantum case) in more than one coordinate
system, then quadratic superintegrability (i.e.,
superintegrability with linear or quadratic constants of motion)
can be considered as a property arising from superseparability.

On the other hand, it is known the existence of some systems endowed
with `higher-order superintegrability', that is, with integrals of
motion which are polynomials in the momenta of order higher than
two (see Refs \cite{GrW02}-\cite{PoPoW12} for some papers
published in these last years). 
Some of them are not separable in any system of coordinates (as the 
Calogero-Moser system \cite{Woj83}--\cite{SmW06}), but others are 
separable in only one particular system of coordinates. 
In this second case, the system possesses two quadratic integrals 
(arising from separability) and a third integral of higher order. 

We will focus our attention on two higher order superintegrable systems 
related with the SW system. 

\begin{itemize}
\item[(i)]  A natural generalization of $V_{sw}$ is given by the following potential
$$
  V(n_x,n_y)  =  {\smallonehalf}{\om_0}^2(n_x^2 x^2 + n_y^2 y^2)
    + \frac{k_1}{2x^2}  + \frac{k_2}{2y^2}  \,,  \label{Van1n2}
$$
that contains a more general  harmonic oscillator with anisotropy \cite{EvVe08}--\cite{RaRoS10}.
When $n_x=n_y=1$ it becomes the SW system, but in the general case it is only separable in Cartesian coordinates; so, the third integral is a polynomial in the momenta of higher order than 2. 

\item[(ii)] The following potential 
$$
  V_{ttw}(r,\phi)  =  {\smallonehalf}\,{\om_0}^2 r^2 +  \frac{1}{2\,r^2}\,\Bigl(   \frac{\alpha}{\cos^2(k\phi)}  +  \frac{\beta}{\sin^2(k\phi)} \Bigr) \,,
  $$
has been firstly  studied by Tremblay, Turbiner, and Winternitz \cite{TTW09}--\cite{TTW10}, and then by other authors  \cite{Qu10a}--\cite{CalCedOlm12}.  
When $k=1$ it becomes the SW system, but in the general $k\ne 1$ case 
it  is only separable in polar coordinates; therefore, the third integral must be a polynomial in the momenta of higher order than two. 

\end{itemize}

The superintegrability of the potential $V(n_x,n_y)$ has been studied using different approaches \cite{EvVe08,RodTW08,RodTW09,RaRoS10}. We point out that it was proved in \cite{RaRoS10} the existence of a complex factorization for the higher-order constant of the motion. 
This approach was in fact a generalization of a method previously used \cite{JauHillPr40,Perelomov} for the superintegrability of the harmonic oscillator 
with commensurable frequencies (the ratio is a rational number).  
Concerning the TTW system it has been studied at both  the classical and the quantum levels.  
We quote here some different approaches: 
(i) Proof that the quantum system is exactly solvable  and that it  is endowed with certain properties that represent a very strong indication of superinegrability \cite{TTW09}; 
(ii) Proof that all bounded classical trajectories are closed  \cite{TTW10}; 
(iii) Techniques related with supersymmetric quantum mechanics \cite{Qu10a}-\cite{Qu10b}; 
(iv) Hamilton-Jacobi formalism \cite{Gon10ArXiv}; 
(iii) As a particular case arising in the general study of the Schr\"odinger equation for quantum systems that admit a separation of variables in some orthogonal coordinate system. \cite{KaKrMi10}. 
We also mention that it has been the starting point for the study \cite{Qu10a}, \cite{Qu10b}, \cite{CalCedOlm12}, of quantum algebraic properties related with  intertwining operators. Very recently, Gonera has proved (in a very modified version of \cite{Gon10ArXiv}) the superintegrability of this system making use of the formalism of the angle-action variables \cite{Gon12PLa}. 

We also mention that Post and Winternitz have studied in \cite{PostWint10} the quantum superintegrability of a family of Hamiltonians with a deformed Kepler-Coulomb potential, and also depending of a parameter $k$, by relating this new family with the TTW system using  the so-called coupling constant  metamorphosis  (or St\"ackel transform) \cite{KaMiPost10} as an approach.

The purpose of this paper is twofold: to study the existence of higher-order superintegrable systems that are separable in polar coordinates  using as an approach the method applied for the study of the superintegrability of $V(n_x,n_y)$ and to study the superintegrability of the  Tremblay-Turbiner-Winternitz system.

 The structure of this article is as follows:
in Sec. 2, we present the main characteristics of the method in the Cartesian case; 
we first consider the linear harmonic oscillator and then we recall the proof of the superintegrability of the nonlinear system $V(n_x,n_y)$.
 In Sec. 3, we study the higher-order superintegrability of a family of potentials separable in polar coordinates and we prove the superintegrability of the TTW system. In Sec. 4, we obtain a new family of superintegrable potentials obtained from the previous one by making use of a transformation (we also study the relation of this family with the TTW system). 
Finally,  in Sec. 5, we make some comments and present some open questions.

\section{Complex factorization and superintegrability : Sepa\-rable systems in Cartesian coordinates }

In this section we recall the method of the complex factorization for two oscillators separable in Cartesian coordinates. 

\subsection{Superintegrability of the linear  oscillator }\label{Sec21}

The two-dimensional  harmonic oscillator
\begin{equation}
  L_{HO} =  {\smallonehalf} (v_x^2 + v_y^2)
  -   {\smallonehalf} ({\om_1}^2 x^2 + {\om_2}^2 y^2)
\end{equation}
has the two one-degree of freedom energies, $I_1=E_x$ and
$I_2=E_y$, as fundamental integrals.
The superintegrability of the rational case,
$\om_1 =n_x {\om_0}$, $\om_2 = n_y {\om_0}$,
with integers $n_x, n_y$, can be proved by making use
of a complex formalism \cite{JauHillPr40,Perelomov}.
Let $A_i$, $i=x,y$, be the following two complex functions
$$
 A_x = v_x + {\ii} n_x {\om_0}\, x \,,{\quad}
 A_y = v_y + {\ii} n_y {\om_0}\, y \,,
$$
then we have the following time-evolution
$$
  \frac{d}{dt}A_x = {\ii} n_x {\om_0}\,A_x \,,{\quad}
  \frac{d}{dt}A_y = {\ii} n_y {\om_0}\,A_y \,.
$$
Thus, the functions $A_{ij}$ defined as
$$
 A_{ij} = (A_i)^{n_y} (A_j^{*})^{n_x} \,,{\quad} i,j=x,y,
$$
are  constants of motion. The two real functions $|A_{xx}|^2$ and
$|A_{yy}|^2$ are proportional to the energies $E_x$ and $E_y$ and
concerning $A_{xy}$, since it is a complex function, it determines
not one but two real first integrals, $\Im (A_{xy})$ and $\Re
(A_{xy})$. So, we have obtained four integrals but, since the system
is two-dimensional, only three of them can be independent. We can
choose  $I_3=\Im (A_{xy})$ as the third fundamental constants of motion 
(the other constant $I_4=\Re(A_{xy})$ can be expressed as a function 
of $E_x$, $E_y$, and $\Im(A_{xy})$).

\subsection{Superintegrability of a generalizated version of the  SW system} \label{Sec22}

A natural generalization of the SW potential \cite{FrMS65, Ev90} is given by 
\begin{equation}
  V(n_x,n_y)  =  {\smallonehalf}{\om_0}^2(n_x^2 x^2 + n_y^2 y^2)
    + \frac{k_1}{2x^2}  + \frac{k_2}{2y^2}  \,,  \label{Van1n2}
\end{equation}  
representing an harmonic oscillator with rational ratio of frequencies and
inversely quadratic nonlinearities.
 It is separable but not multiple separable and because of this the additional constants of the motion are of higher order in the momenta.   It has been proved in \cite{RaRoS10} that it also admits a complex factorization that was obtained as a deformation of the quadratic version of the factorization of the integrals of motion of the linear oscillator.

The proof is as follows.  Let  us now denote by $B_i$, $i=x,y$, the following two complex functions
$$
 B_{x} = \Bigl(v_x^2 - n_x^2 \om_0^2 x^2 +\frac{k_1}{x^2}\Bigr) + 2\,{\ii} n_x {\om_0} x v_x\,,{\quad}
 B_{y} = \Bigl(v_y^2 - n_y^2 \om_0^2 y^2 +\frac{k_2}{y^2}\Bigr) + 2\,{\ii} n_y {\om_0} y v_y\,.
$$
then we have the following time-evolution
$$
 \frac{d}{d t}\,B_x = 2\,{\ii} n_x {\om_0}\,B_x  \,,{\quad}
 \frac{d}{d t}\,B_y = 2\,{\ii} n_y {\om_0}\,B_y  \,.
$$
Thus, the complex functions $B_{ij}$ defined as
$$
  B_{ij} = (B_i)^{n_j}\,(B_j^{*})^{n_i}\,,{\quad} i,j=x,y,
$$
are constants of the motion
$$
  \frac{d}{dt}\,B_{xy}  =   B_x^{(n_y-1)}B_y^{(n_x-1)}\,\Bigl( \,   n_y\dot{B_x}\,B_y^{*}
   +  n_x B_x\,\dot{B}_y^{*} \,\Bigr)  =  0  \,.
$$
The moduli of $B_x$ and $B_y$ are related with the energies 
$$
 |B_x|^2 = 4 (E_x^2 - k_1 n_x^2 \om_0^2) \,,{\quad}
 |B_y|^2 = 4 (E_y^2 - k_2 n_y^2 \om_0^2) \,.
$$
and, as in the linear case, $B_{xy}$ can be considered as coupling the 
two degrees of freedom.

\section{Complex factorization and superintegrability : Sepa\-rable systems in polar coordinates }

The two systems studied in Secs. (\ref{Sec21}) and  (\ref{Sec22}) were separable in Cartesian coordinates and the two functions, $A_i$ in the linear case and $B_i$ in the nonlinear case, were dependent of only one coordinate. Now, in this section, we study the existence of a complex factorization but for potentials separable in polar coordinates $(r,\phi)$. The main difference is that now we present the inverse approach; we start with a general separable potential $V(r,\phi)$ and then we determine the conditions to be satisfied by the potential to be superintegrable with the additional constant of motion given by the product of two complex functions. 

We will make use of the Hamiltonian formalism, so, in what follows, the time derivative $d/dt$ means Poisson bracket with the function $H$.

Let us consider the following Hamiltonian 
\begin{equation}
 H =   {\smallonehalf}\,\bigl(p_r^2 + \frac{p_\phi^2}{r^2}\bigr) + 
 {\smallonehalf}\,{\om_0}^2 r^2 + {\smallonehalf}\frac{F(\phi)}{r^2} \,.  \label{H(rfi}
\end{equation}
It is separable in polar coordinates and  it is therefore endowed with the following  two constants of the motion 
\begin{eqnarray*}  
 J_1 &=& p_r^2 + \frac{p_\phi^2}{r^2} + {\om_0}^2 r^2 +\frac{F(\phi)}{r^2}  \cr 
 J_2 &=& p_\phi^2 + F(\phi)
\end{eqnarray*}
Any property to be satisfied by a Hamiltonian as (\ref{H(rfi}) must be true for the case of the single oscillator (without additional $\phi$-dependent term). So we first consider the case of a central potential without the function $F$.

{\sl Step 1.}
Let us first consider a central potential $U(r)$. Then if we denote by $M_r$ and $M_i$ the functions
$$
 M_r =  \frac{2}{r}\,p_r\,p_\phi \,,{\quad}
 M_i =  p_r^2 -  \frac{p_\phi ^2}{r^2} + U(r) \,,
$$
then we have 
\begin{itemize}
\item[(i)] The time-derivative of $M_i$ is given by 
$$
 \frac{d}{d t}\,M_i = 2\Bigl(\frac{1}{r^2}\,p_\phi \Bigr)M_r  \,. 
$$
\item[(ii)] The time-derivative of $M_r$ is given by 
$$
 \frac{d}{d t}\,M_r =   -2\Bigl(\frac{1}{r^2}\,p_\phi \Bigr)M_i \,,  
$$
if and only if the potential $U$ is solution of the following equation 
$$
 r \,U' - 2 U = 0 \,. 
$$
\end{itemize}

The harmonic oscillator satisfies the properties (i) and (ii) with the factor  $\lambda_0$ given by $\lambda_0 = (p_\phi/r^2)$.   It seems natural to assume that, when considering the more general Hamiltonian (\ref{H(rfi}), then the momentum $p_\phi$  in $M_r$ and  $\lambda_0$  will probably change to $\sqrt{J_2}$. 

{\sl Step 2.}
Consider now the non-linear potential
$$
  V(r,\phi)  =  {\smallonehalf}\,{\om_0}^2 r^2 + {\smallonehalf}\frac{F(\phi)}{r^2}  \,.  \label{Vr2Ffi}
$$
Then if we denote by $M_r$ and $M_i$ the functions
$$
 M_r =  \frac{2}{r}\,p_r\,\sqrt{J_2} \,,{\quad}
 M_i =   J_1 - \frac{2}{r^2} \, J_2  \,. 
$$
the time-evolution of them is given by
$$
 \frac{d}{d t}\,M_r =  2 (-{\lambda})\,M_i  \,,{\quad}
 \frac{d}{d t}\,M_i =  2 {\lambda}\,M_r  \,,{\quad} {\lambda} = \frac{1}{r^2}\,\sqrt{J_2}  \,. 
$$

{\sl Step 3.}
Consider a potential separable in polar coordinates. 
$$
  V(r,\phi)  = U(r) + {\smallonehalf}\frac{F(\phi)}{r^2}    \label{Vr2Ffi}
$$
and denote by $N_r$  the following function
$$
 N_r =  z +  J_2\,\cos(k\phi)  \,,{\quad}
$$
where $z$ denotes a real parameter. Then we have 
\begin{itemize}
\item[(i)] The time-derivative of $N_r$ is given by 
$$
 \frac{d}{d t}\,N_r =  (-{\lambda})\,k\,N_i  \,,{\quad} N_i =   \sqrt{J_2} \,p_\phi\,\sin(k\phi)\,,{\quad}
 {\lambda} = \frac{1}{r^2}\,\sqrt{J_2} \,.  
$$
\item[(ii)] The time-derivative of $N_i$ is given by 
$$
 \frac{d}{d t}\,N_i =  {\lambda}\,k\,N_r \,,
$$
if and only if the function $F$ is solution of the following equation 
$$
 \sin(k\phi) \,F' + 2 k \cos(k\phi)\,F + 2 k z = 0 \,. 
$$
\end{itemize}
This is a linear first order equation for $F$ and the general solution is given by 
\begin{eqnarray*}  
 F = \frac{k_a} {\sin^2(k\phi)} +  2 z\,\Bigl(\frac {\cos(k\phi)} {\sin^2(k\phi)}\Bigr) \,. 
\end{eqnarray*}  
Notice that $F$ is a linear combination of two functions because the presence of the parameter $z$. If $z=0$ then $F$ reduces to the solution of the homogeneous version of the linear equation.

The steps 1 and 2 provide a property that singles out  the harmonic oscillator in the set of all central potentials; on the other hand, the step 3 states the existence of an appropriate function $N$ for any function $U(r)$ and certain  very particular values of the function $F(\phi)$.    The superposition of these two properties leads to the following proposition. 

\begin{proposicion}  \label{prop1}
Consider the non-linear potential
\begin{equation}
  V_1(r,\phi)  =  {\smallonehalf}\,{\om_0}^2 r^2 + {\smallonehalf}\frac{F_1(\phi)}{r^2}   \,,{\quad}
  F_1 = \frac{k_a} {\sin^2(k\phi)} +  k_b\,\Bigl(\frac {\cos(k\phi)} {\sin^2(k\phi)}\Bigr) \,,
\end{equation}
where $k_a$ and $k_b$ are arbitrary constants.
Let $M$ and $N$ be the  complex functions   $M = M_r + i\,M_i$ and  $N = N_r + i\,N_i$ with real and imaginary parts $M_a$ and $N_a$, $a=r,i$, be defined as 
$$
 M_r =  \frac{2}{r}\,p_r\,\sqrt{J_2} \,,{\quad}
 M_i =   J_1 - \frac{2}{r^2} \, J_2  \,,
$$
$$
 N_r =  \frac{k_b}{2} +  J_2\,\cos(k\phi)  \,,{\quad}
 N_i =   \sqrt{J_2} \,p_\phi\,\sin(k\phi) \,.
$$ 
Then, the complex function $K$ defined as
$$
  K = M^{k} (N^{*})^{2}
$$
is a (complex) constant of the motion.
\end{proposicion}

{\it Proof:} 
First, let us comment that if $k$ is an integer $(k=n)$ then $V_1$ is well defined as a function on the plain and the domain (configuration space of the system) is the union of  $2n$ wedges that cover the plain (the singularities of $V_1$ are in the borderlines).  If $k$ is a rational number  then the configuration space must be restricted to only one wedge of angle smaller than $2\pi$.

The time-evolution of the functions $M$ and $N$ is given by
$$
 \frac{d}{d t}\,M  = {\ii} 2 {\lambda}\,M  \,,{\quad}
 \frac{d}{d t}\,N  = {\ii} k\,{\lambda}\,N   \,,{\quad}
  {\lambda} = \frac{1}{r^2}\,\sqrt{J_2}\,. 
$$
Thus we have
\begin{eqnarray*}
  \frac{d}{dt}\,K &=&  M^{(k-1)}N^{*}\,\Bigl( \, k\dot{M}\,N^{*}
   +  2 M\,\dot{N}^{*} \,\Bigr)  =  0   \,.
\end{eqnarray*}
\hfill${\square}$

Next, we give the expressions of the function $F_1$ for the four first integer values of $k$:
\begin{eqnarray*}
 F_1(k=1) &=& \frac{k_a}{y^2} + \frac{k_k\,x}{y^2\sqrt{x^2+y^2}}  \cr
 F_1(k=2) &=& \frac{k_a-k_b}{4x^2} + \frac{k_a+k_b}{4y^2}  \cr  
 F_1(k=3) &=& \frac{1}{(3x^2-y^2)^2y^2} \Bigl( k_a (x^2+y^2)^2 
 + k_b (x^2-3y^2) x \sqrt{x^2+y^2}  \Bigr) \cr 
 F_1(k=4) &=&  \frac{1}{16(x^2-y^2)^2x^2y^2} \Bigl(
 k_a (x^2+y^2)^3 + k_b(x^2+y^2)(x^4 - 6 x^2 y^2 + y^4)  \Bigr) 
\end{eqnarray*}

At this point we recall that the potential of Tremblay-Turbiner-Winternitz \cite{TTW09}--\cite{CalCedOlm12} is given by 
$$
  V_{ttw}(r,\phi)  =  {\smallonehalf}\,{\om_0}^2 r^2 +  \frac{1}{2\,r^2}\,\Bigl(   
  \frac{\alpha} {\cos^2(k\phi)} +  \frac {\beta} {\sin^2(k\phi)} \Bigr) \,. 
$$
It can be proved that the function $F_1$ reduces to the $\phi$-dependent part of the TTW potential when (i) the coefficients $k_a$ and $k_b$ are writen as $k_a=2(\alpha+\beta)$ and $k_b=2(\beta -\alpha)$, and (ii) the angular coefficient $k$ is changed to $2k$; that is,  
$$
  \frac{2(\alpha+\beta)} {\sin^2(2k\phi)} +  2(\beta - \alpha)\,\Bigl(\frac {\cos(2 k\phi)} {\sin^2(2k\phi)}\Bigr) =  \frac{\alpha} {\cos^2(k\phi)} +  \frac {\beta} {\sin^2(k\phi)}  \,, 
$$
and consequently 
$$
  V_1(r,\phi,2k)  =  V_{ttw}(r,\phi,k)  \,. 
$$
In fact,  we must point out that the expression obtained for $K$ in proposition  \ref{prop1}  is quite similar (introducing the appropriate changes of coefficients and notatios) to the expression obtained by Gonera in \cite{Gon12PLa} making use of the action-angle formalism. 

Summarizing, the above proposition  \ref{prop1} states that the potential $V_1$, that is a separable potential, it is also superintegrable with a third constant of the motion that arises as the real or the imaginary part of a complex function $K$ obtained from the functions $M$ and $N$. 
Moreover, this potential is equivalent to the TTV potential; therefore, proposition  \ref{prop1} also proves the superintegrability of the TTW system.

\section{A second family of superintegrable systems }

Suppose that we interchange the  $\cos(k\phi)$ and the $\sin(k\phi)$ functions in the expression of the function $N$.  
In this case we  then we arrive (in the step 3)  to the equation 
$$
 \cos(k\phi) \,F' + 2 k \sin(k\phi)\,F +  2 k z = 0 \,, 
$$
wich has as general solution the function $F_2$ given by  
$$
  F_2 = \frac{k_a} {\cos^2(k\phi)} +  k_b\,\Bigl(\frac {\sin(k\phi)} {\cos^2(k\phi)}\Bigr)
 {\quad}(k_b=2 z) \,. 
$$

Consequently, we can also state that the potential
\begin{equation}
  V_2(r,\phi)  =  {\smallonehalf}\,{\om_0}^2 r^2 + {\smallonehalf}\frac{F_2(\phi)}{r^2}   \,,{\quad}
  F_2 = \frac{k_a} {\cos^2(k\phi)} +  k_b\,\Bigl(\frac {\sin(k\phi)} {\cos^2(k\phi)}\Bigr) \,,
\end{equation}
is superintegrable with a third (complex) constant of the motion given by 
$$
  K = M^{k} (N_2^{*})^{2}  \,, 
$$
where $M$ is defined as in proposition 1 and $N_2$ is now defined as$N_2 = N_{2r} + i\,N_{2i}$ with 
 $$
 N_{2r} =  \frac{k_b}{2} +  J_2\,\cos(k\phi)  \,,{\quad}
 N_{2i}=   \sqrt{J_2} \,p_\phi\,\sin(k\phi) \,.
$$

Next, we give the expressions of the function $F_2$ for the four first integer values of $k$:
\begin{eqnarray*}
F_2(k=1) &=& \frac{k_a}{x^2} + \frac{k_k\,y}{x^2\sqrt{x^2+y^2}}  \cr
F_2(k=2) &=& k_a\,\frac{x^2+y^2}{(x^2-y^2)^2} +  k_b\,\frac{2 x y}{(x^2-y^2)^2}  \cr  
F_2(k=3) &=& \frac{1}{(x^2-3y^2)^2x^2} \Bigl( k_a (x^2+y^2)^2 
+ k_b (3x^2-y^2) y \sqrt{x^2+y^2}  \Bigr) \cr 
F_2(k=4) &=&  \frac{1}{(x^4 - 6 x^2 y^2 + y^4)^2} \Bigl(
k_a (x^2+y^2)^3 + 4 k_b(x^4 - y^4) xy \Bigr) 
\end{eqnarray*}

Finally,  we consider the relation of $F_2$ with $F_1$ and with TTW.

\begin{itemize}
\item[(i)]   The interchange the  $\cos(k\phi)$ and the $\sin(k\phi)$ functions can also be seen as a rotation of angle $\pi/(2 k)$. That is, 
$$
  F_2\bigl(\phi \to \phi + \pi/(2 k)\bigr) = F_1  \,. 
$$
The kinetic term in $H$ is rotationally invariant; so any rotation (arbitrary value of the  angle) transforms the potential $V_1$ in a new potential that is therefore also integrable (the new constants are the rotated versions of the original constants).  From this viewpoint, the potential  $V_2$ can be seen not just as a new potential but as a rotated version of $V_1$. Nevertheless  the transformation $V_1\to V_2$ is rather peculiar and interesting since it is a  different rotation for every value of $k$ and is such that it preserves the high degree of simplicity of the potential. 
We also note that the domain of $V_2$ is different to the domain of $V_1$.

\item[(ii)]   We have proved that 
$$
  V_1(r,\phi,k)  =  V_{ttw}(r,\phi,k/2) =  {\smallonehalf}\,{\om_0}^2 r^2 +  \frac{1}{2\,r^2}\,\Bigl(  \frac{2\alpha} {1+\cos(k\phi)} +  \frac {2\beta} {1-\cos(k\phi)} \Bigr) \,. 
$$
Then if we replace $\cos(k\phi)$ by $\sin(k\phi)$, we obtain the following TTW-related potential 
$$
 \wt{V}_{ttw}(r,\phi,k)    = {\smallonehalf}\,{\om_0}^2 r^2 +  \frac{1}{2\,r^2}\,\Bigl(  \frac{2\alpha} {1+\sin(k\phi)} +  \frac {2\beta} {1-\sin(k\phi)} \Bigr) \,. 
$$
So that we arrive at the following relation
$$
  V_2(r,\phi,k)  =  \wt{V}_{ttw}(r,\phi,k)  \,. 
$$

\end{itemize}

\section{Comments and open questions  }

We have studied the existence of systems that are separable in polar coordinates and admit  an additional constant of motion of higher order in the momenta. The main idea is that this additional constant can be factorized as the product of powers of two particular rather simple complex functions  (here denoted by $M$ and $N$).  An important point is that all the potentials obtained by this method are related to the harmonic oscillator. As is well known, the other fundamental superintegrable central potential is the potential of the Kepler problem (this 
is just the theorem of Bertrand).  The Kepler problem admits two noncentral superintegrable potentials separable in two different systems (this was proved in \cite{FrMS65}).  Moreover, the existence of deformations with higher order superintegrability has been proven recently \cite{PostWint10}.  It would be  convenient try to apply the procedures presented in this study to the Kepler problem.  This  question, as well as some other related problems (as the properties of the quantum version of the functions $M$ and $N$),  is an open question to be studied.

\section*{Acknowledgments}

This work was supported by the research projects MTM--2009--11154 (MEC, Madrid)  and DGA-E24/1 (DGA, Zaragoza).

{\small

\end{document}